\documentclass[prl,showpacs,preprintnumbers,amsmath,amssymb,letterpaper,twocolumn]{revtex4}

\usepackage{graphicx}
\usepackage{dcolumn}
\usepackage{bm}
\usepackage{epstopdf}
\usepackage{latexsym}
\usepackage{epsfig}
\usepackage{verbatim}
\usepackage{hyperref}
\usepackage{color}

\newcommand{\be}{\begin{equation}}
\newcommand{\ee}{\end{equation}}
\newcommand{\bea}{\begin{eqnarray}}
\newcommand{\eea}{\end{eqnarray}}

\newcommand{\eref}[1]{Eq.~(\ref{#1})}

\begin{document}

\title{Spin-Resolved Tunneling Studies of the Exchange Field in EuS/Al Bilayers}

\author{Y.M. Xiong, S. Stadler, P.W. Adams}
\affiliation{Department of Physics and Astronomy, Louisiana State
University, Baton Rouge,
Louisiana 70803, USA}
\author{G. Catelani}
\affiliation{Department of Physics, Yale University, New Haven,
Connecticut 06520, USA}

\date{\today}

\begin{abstract}
We use spin-resolved electron tunneling to study the exchange field
in the Al component of EuS/Al bilayers, in both the superconducting and normal-state phases of the Al.
Contrary to expectation, we show that the exchange field, $H_{ex}$, is a non-linear function of applied field, even in
applied fields that are well beyond the EuS coercive field.  Furthermore the magnitude $H_{ex}$ is unaffected by the superconducting phase.  In
addition, $H_{ex}$ decreases significantly with increasing
temperature in the temperature range of 0.1 - 1 K.  We discuss these results in the context of
recent theories of generalized spin-dependent boundary conditions at
a superconductor/ferromagnet interface.
\end{abstract}


\maketitle

Owing to their different symmetries, itinerant ferromagnetic
(FM) order and spin-singlet superconducting (SC) order are generally mutually
exclusive.  With rare exception, nature does not allow ferromagnetic
order to coexist with BCS superconductivity \cite{UGe2}.  This immiscibility can, however, lead to interesting effects in
the vicinity of a FM/SC interface, as electrons moving from one region to the other try to
accommodate the differing order parameters.  Over the last decade
significant progress has been made in understanding the nature of
the SC order parameter in the proximity of a FM/SC interface \cite{Buzdin1,Beasley,Efetov,Kontos}.  In
fact, much of the research on FM/SC structures has focused on the
evanescent SC condensate residing on the FM side of the
interface in properly prepared bilayers \cite{Efetov,Cottet1,Cottet2,Cottet3}.  Remarkably, not only can
SC Cooper pairs exist in the FM layer, but the exchange field in the
FM induces a triplet component to the SC wavefunction \cite{Efetov}.  This results in oscillations in the SC order parameter \cite{Kontos,Cottet1,Cottet2} that are
reminiscent of the order parameter modulations predicted by Fulde
and Ferrel \cite{FF}, and Larkin and Ovchinnikov \cite{LO} for a BCS
superconductor in a critical Zeeman field.  In the present Letter,
we present the results of a detailed study of the exchange field
induced in the {\it SC side} of a FM/SC bilayer.  We show this
proximity-induced exchange field is not static, but has unexpected
temperature and applied-field dependencies that are not attributable
to the temperature and/or field dependence of the FM magnetization.

Since we are primarily interested in the behavior of the exchange
field induced in the SC layer, we have chosen an insulating material
for the FM layer, EuS \cite{EuS}.  This greatly simplifies the interpretation
of the data since electrons from the SC only enter the FM via
evanescent wavefunctions.  For the superconductor we chose Al since
it has a very low spin-orbit scattering rate and its
spin-paramagnetic phase diagram is well understood \cite{Adams1}.   The FM/SC
bilayers were fabricated by first depositing a 5 nm-thick EuS film
via e-beam evaporation onto fire-polished glass at 84~K.  Then a 2.4 nm thick Al
film was deposited on top of the EuS film.  The depositions were
made at a rate of $\sim1$~nm/s and $\sim0.1$~nm/s, respectively, in a typical vacuum of
$<3\times10^{-7}$ Torr.  The samples
were then exposed to air to form a native oxide on the Al surface.  Finally, the
bilayers where mechanically trimmed and a 10 nm-thick
non-superconducting Al alloy counter-electrode (Al$_{ns}$) was deposited, with the
native oxide serving as the tunnel barrier.  The
junction area was about 1~mm$\times$1~mm, while the junction
resistance ranged from 15-100~k$\Omega$ depending on exposure time
and other factors.  Only junctions with resistances much greater
than that of the films were used.  At low temperature the tunneling
conductance is proportional to the quasiparticle density of states (DOS) \cite{Tinkham}.
Measurements of resistance and tunneling were carried out on an
Oxford dilution refrigerator using a standard ac four-probe
technique. Magnetic fields of up to 9~T were applied using a
superconducting solenoid. A mechanical rotator was employed to
orient the sample \textit{in situ} with a precision of
$\sim0.1^{\circ}$.

The exchange field in both the SC and normal phases of the bilayers
can be obtained via spin-resolved tunneling DOS measurements.  For
samples in the SC phase, we utilize the quasiparticle tunneling
technique of Tedrow and Meservey \cite{TMreview}. This technique exploits
the fact that the internal field in the SC film induces a Zeeman
splitting of the DOS spectrum, resulting in the BCS coherence peaks
being split into spin-up and spin-down bands.  The separation of the
bands, $\Delta V=E_z/e$ (with $E_z=2\mu_B H_z$ the Zeeman energy),
is a direct measure of the effective Zeeman field
\be
H_{z}=\frac{H_{app}+H_{ex}}{1+G^0_\mathrm{eff}},
\label{Spectra}
\ee
where $H_{app}$ is the applied field, $H_{ex}$ the  exchange field
induced by the EuS interface, and $G^0_\mathrm{eff}$ is the
effective~\cite{CWA} antisymmetric Fermi liquid parameter. The
latter accounts for the renormalization of the electron spin by
interactions.  At low temperatures where $T\ll T_c$ $G^0_\mathrm{eff} \simeq 0$, whereas in the normal-state
$G^0_\mathrm{eff}=G^0_N$ with $G^0_N\simeq 0.16-0.26$ \cite{CWA}.

In the upper panel of Fig.~\ref{Spectra} we plot the 80~mK tunneling
conductance of an EuS/Al-AlO$_x$-Al$_{ns}$ tunnel junction in an
applied parallel magnetic field of $H_{app}=0.03$~T.  The Zeeman
splitting of the BCS DOS spectrum is clearly evident.  From this
splitting we obtain $H_{z}\simeq 4.4$~T.  When the Zeeman energy is
of the order of the superconducting gap $\Delta_o\sim0.4$~mV the
film undergoes a first-order transition to the normal state at the
analog of the Clogston-Chandrasekhar critical field \cite{CC}, \be
H_{z}^{c}= \frac{\Delta_o}{\mu_B\sqrt{2(1+G^0_N)}}.
\label{CriticalField} \ee The parallel critical field in pristine Al
films of comparable thickness to ones used in this study is
$H_z^c\sim6$~T \cite{Adams1}.  For the film in Fig.~\ref{Spectra}
the transition to the normal-state occurred at an applied field of
only $\sim0.1$~T.  This is consistent with the tunneling data and
indicates that the Zeeman field of the Al film was dominated by
exchange.  The magnitude of $H_{ex}$ obtained from data such as that
in Fig.~\ref{Spectra} is comparable to that reported by Hao {\it et
al.} who measured the magnitude of the exchange field as a function
of Al thickness in the SC phase of EuS/Al bilayers \cite{Hao}. Since
the technique used in those early experiments required that the
films be in the SC phase, the exchange fields could only be measured
over a very narrow range of applied fields.  As we discuss in detail
below, by extending the measurements into the normal phase of the
bilayers we are able to measure the exchange field over a much
broader range of applied fields.

\begin{figure}
\begin{flushleft}
\includegraphics[width=.44\textwidth]{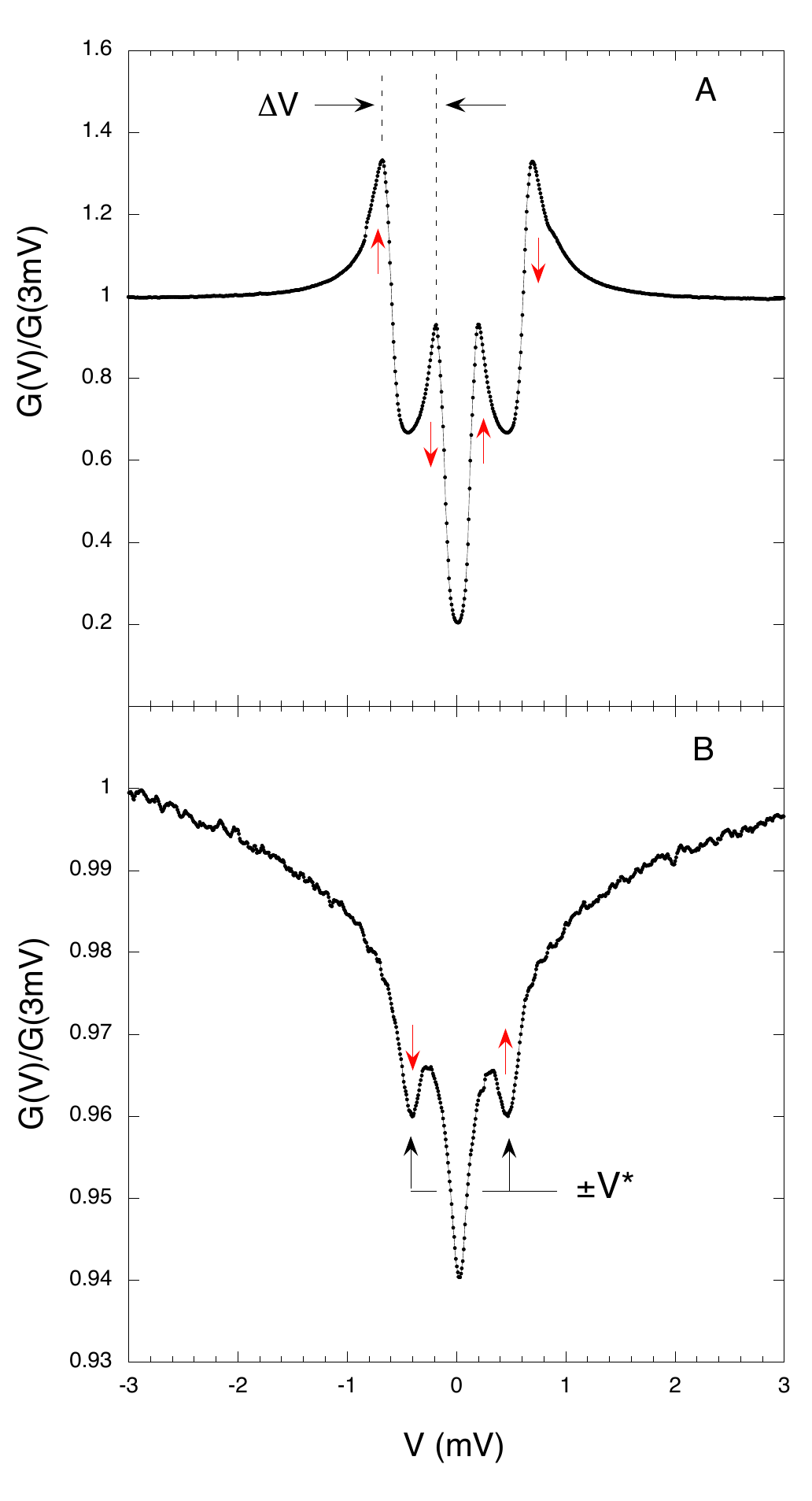}\end{flushleft}
\caption{\label{DoS}  A: Tunneling DOS spectrum in an applied
parallel field of 0.03 T. The Zeeman splitting of the coherence
peaks gives a direct measure of the Zeeman field.  The red arrows
denote the spin assignment of the coherence peaks. B: Pauli-limited
normal state in an applied parallel field of 0.1 T, where only the
pairing resonance and the zero bias anomaly remain.  The Zeeman
field can be extracted from the resonance energy, $\rm V^*$, via
\eref{PR}.  The red arrows denote the spin assignments of the
occupied and unoccupied resonances. }
\end{figure}

In the lower panel of Fig.~\ref{DoS} we plot the tunneling
spectra of the same bilayer in an applied field of $H_{app}=0.1$~T,
which produced a Zeeman field exceeding $H_{z}^{c}$ .  The
central dip in this normal-state spectrum is the {\it electron-electron}
interaction zero-bias anomaly, which is independent of field \cite{ZBA1,ZBA2}.  The
satellite features represent the pairing resonance (PR), from which
we can extract the Zeeman field \cite{PR1,PR2}.  The PR is spin-assigned as shown
by the arrows in the figure.  The energy of the resonance depends on the field
via the Zeeman energy~\cite{PR1},
\be
e{\rm V^*} = \frac{1}{2}\left(E_z +\sqrt{E_z^2-\Delta_0^2}  \right) \ .
\label{PR}
\ee
The positions of the resonances, as shown in
the lower panel of Fig.~\ref{DoS}, are obtained by first
subtracting off the zero-bias anomaly background and then fitting
the resonance profile, as described elsewhere \cite{PR1}.  We then use
\eref{PR} to extract the Zeeman field.

\begin{figure}
\includegraphics[width=.46\textwidth]{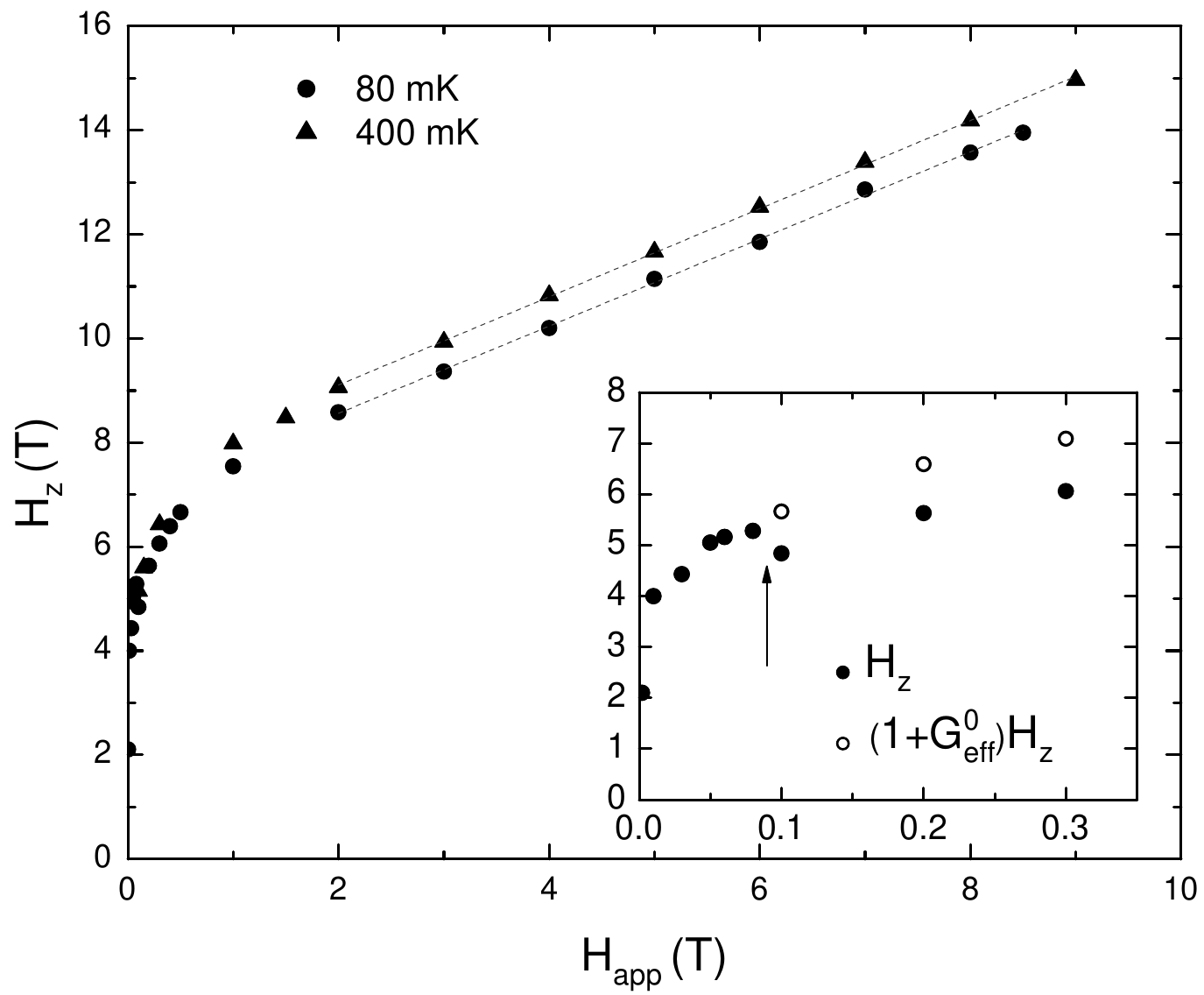}
\caption{\label{Hint}Zeeman field [\eref{Spectra}] as a function
of applied field. The dashed lines represent linear fits to the data
above 2 T.}
\end{figure}

In Fig.~\ref{Hint} we plot $H_{z}$ as a function
of a parallel field $H_{app}$ at 80~mK and 400~mK for
two different samples made under identical
conditions.  The 80~mK data set was obtained from both SC and
normal-state tunneling spectra. This particular sample underwent a first-order
transition to the normal-state at an applied field of $\sim 0.1$~T.
All of the 400~mK data points were obtained from normal-state
spectra, since for this film $H_{ex}>H_z^c$ in zero applied field.  There are several noteworthy features of this data.  The
first is that rather high internal fields can be achieved by
applying fields of a few hundred Gauss.  The second is
that there is a non-linear increase in $H_{z}$ between 0
and 2~T. We ascribe this behavior to a non-linear dependence of the
exchange field $H_{ex}$ on $H_{app}$, as discussed below.
Finally $H_{z}$ increases linearly in applied
fields above 2~T, indicating a saturation of $H_{ex}$ at high fields. The slope is determined by the Fermi
liquid parameter $G^0_N$ as in \eref{Spectra}. We can therefore obtain $G^0_N$
by fitting the data above 2~T to straight lines, as
shown in Fig.~\ref{Hint}. We find $G^0_N=0.19$, $0.18$ for the 80~mK and 400~mK data sets, respectively,
in good agreement with previously measured values in Al films \cite{PR1}.

As a check of the consistency of the above analysis, we show in the inset of Fig.~\ref{Hint} the Zeeman field
vs. $H_{app}$ at 80~mK for low applied fields. The arrow
points to the discontinuity in the Zeeman field at the critical field $H_{z}^c$. The discontinuity is
caused by the jump of $G^0_\mathrm{eff}$ from its SC value ($\simeq 0$) to its normal-state one. Indeed,
$H_z$ multiplied by $1+G^0_\mathrm{eff}$ evolves smoothly with applied field.

\begin{figure}
\begin{flushleft}
\includegraphics[width=.48\textwidth]{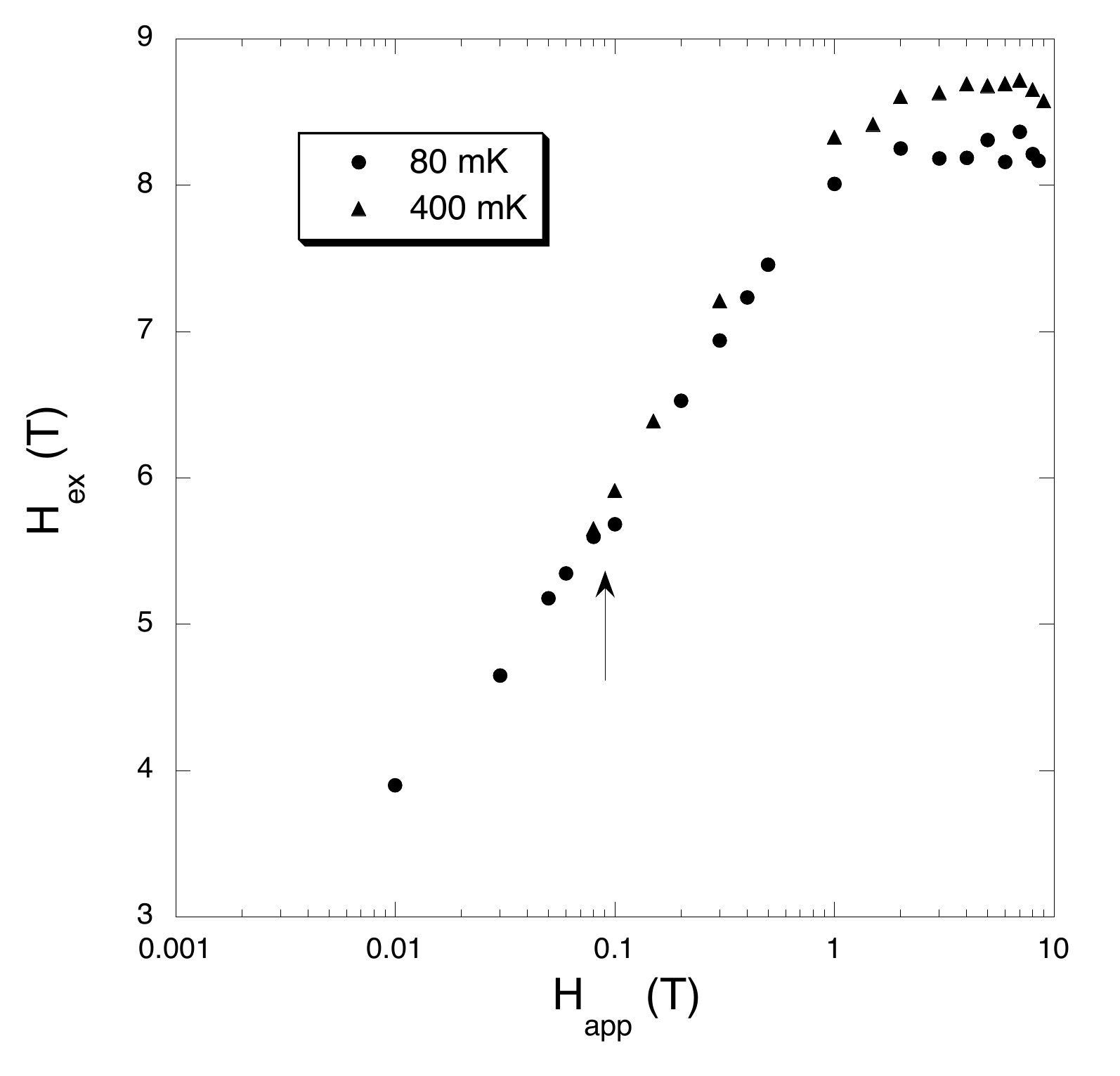}\end{flushleft}
\caption{\label{Hex}Semi-log plot of the EuS exchange field as a function of parallel applied field.
The arrow denotes the superconducting critical field for the 80 mK data.  All of the 400 mK were obtained
from normal-state spectra.}
\end{figure}

The exchange field can be extracted from the data in Fig.~\ref{Hint} by inverting \eref{Spectra}.
In Fig.~\ref{Hex} we show the resulting $H_{ex}$ as a function of
applied field at 80~mK and 400~mK.
The arrow depicts the critical field transition in the 80~mK data
set. Note that, below 2 T, $H_{ex}$ grows non-linearly with applied field, appearing to
increase logarithmically by a factor of 2 between 0.01 T and 1 T.  Also there is no obvious discontinuity
in $H_{ex}$ at the first-order transition.
Similar enhancements in the exchange field with applied field were reported
in both EuO/Al \cite{TTK} and EuS/Al \cite{Hao} bilayers.  (Those measurements, however, were limited to the SC state, while here
we are presenting results for the normal state as well.)
This nonlinear behavior was attributed to the alignment of the ferromagnet domains by
the applied field \cite{TMreview}.  The authors argued that if the FM domains are randomly
oriented on length scales on the order of the superconducting
coherence length, then the average exchange field, as experienced by the SC,
is lowered.  In this scenario the applied field simply serves to align the domains.
In order to explore this as a mechanism for
the behavior in Fig.~\ref{Hex} we have directly measured the
magnetization of the EuS/Al bilayers using a Quantum Design MPMS
SQUID magnetometer.

\begin{figure}
\begin{flushleft}
\includegraphics[width=.48\textwidth]{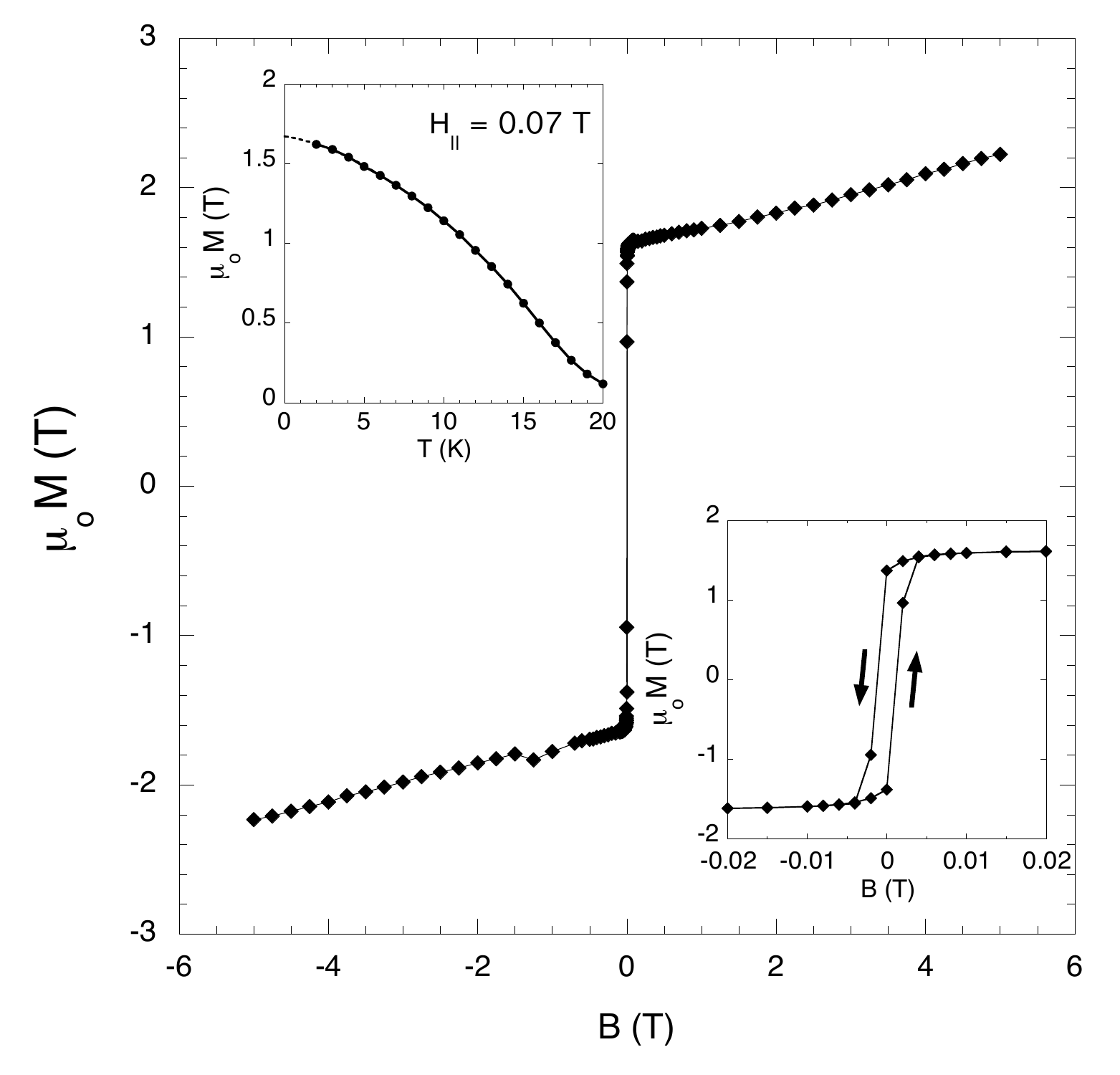}\end{flushleft}
\caption{\label{EuS}Magnetization of a 5 nm-thick EuS film capped with 3 nm of Al.  The external field was applied in the film plane.  Lower inset is an expanded plot of the hysteresis loop.  Note that the coercive field is less than 0.01 T.  The upper inset is the temperature dependence of the magnetization in a 0.07 T parallel field.  This data was taken after cycling the applied field to 5 T and back at 2 K.  The dashed line in this inset is a polynomial fit to the data below 5 K.}
\end{figure}

\begin{figure}
\begin{flushleft}
\includegraphics[width=.48\textwidth]{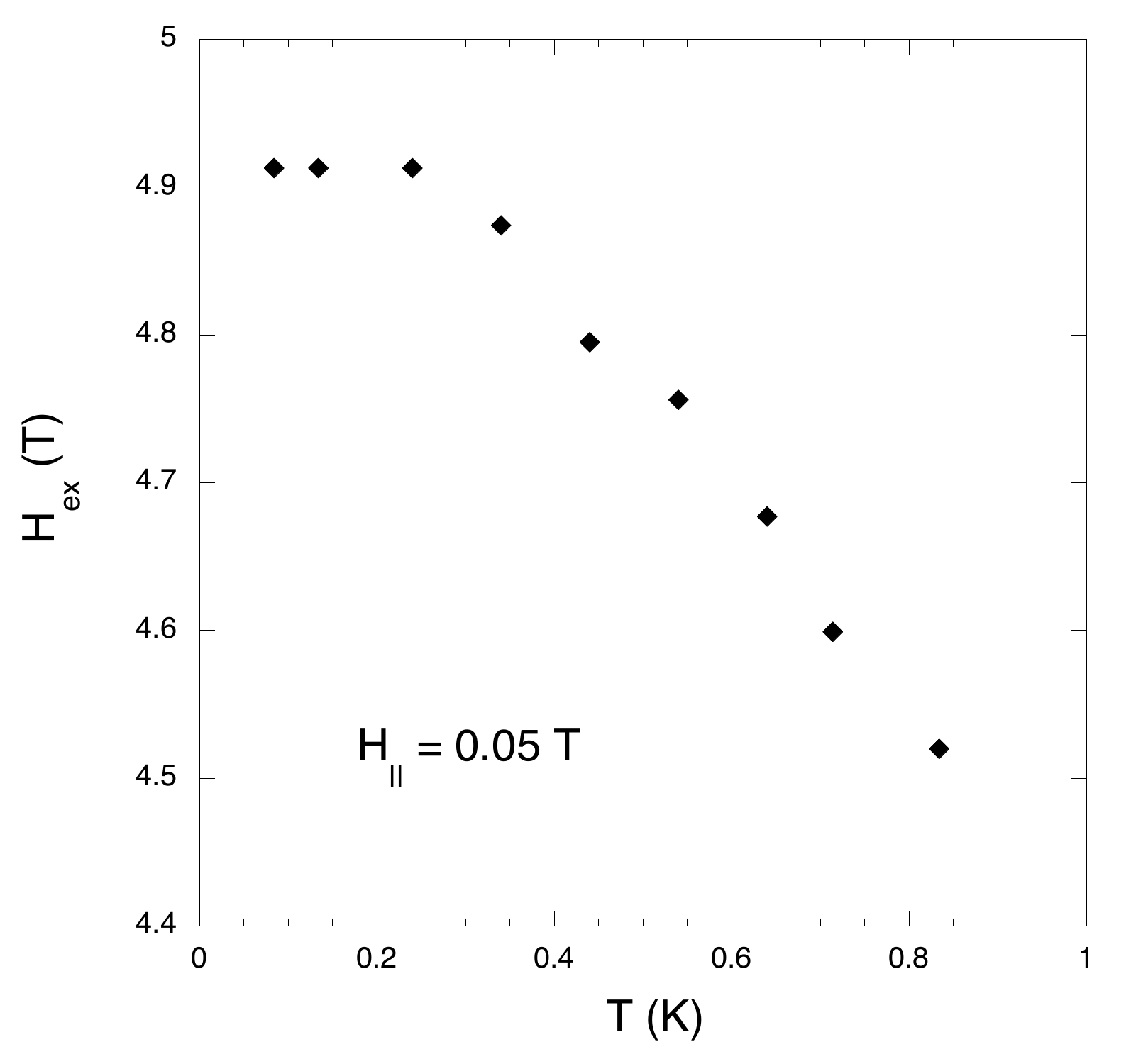}\end{flushleft}
\caption{\label{Hex-T}Temperature dependence of the exchange field in a parallel applied field of 0.05 T.}
\end{figure}

In Fig.~\ref{EuS} we show the longitudinal magnetization, with field
oriented along the film plane, of a stack of 10 EuS/Al bilayers at
2~K.  The background magnetization of the glass slides was measured
separately and subtracted from the raw data.  Note that the
magnetization loop is very sharp with a coercive field below 100~G.
The saturation magnetization and Curie temperature are in good agreement
with those of bulk EuS \cite{EuS}.  We find no evidence that the ferromagnetic behavior of the EuS has been
significantly affected by its contact with the Al layer, as was conjectured in
Ref.~\onlinecite{Tokuyasu}. This suggests
that observed increase in $H_{ex}$ in applied fields between 0.01~T
and 2~T is not an artifact of domain alignment but is, in fact, an
intrinsic effect.  Interestingly, $H_{ex}$ also exhibits a
significant temperature dependence below 1~K.  In Fig.~\ref{Hex-T}
we plot the exchange field as a function of temperature in the
presence of a parallel applied-field of $H_{app}=0.05$~T.  Note that
the $H_{ex}$ decreases by about 10\% between 200 and 800~mK.  The magnetization of the EuS below 2 K (see upper inset of Fig.~\ref{EuS}) is only weakly temperature dependent and cannot explain the decrease $H_{ex}$ with increasing temperature.  This suggest that the behavior in Fig.~\ref{Hex-T} is a conduction-spin relaxation effect associated with thermally activated spin-flip scattering processes.

In the diffusive regime relevant to our Al films, the spin-dependent
boundary conditions described in Refs.~\onlinecite{Cottet2} and \onlinecite{Cottet3} should produce an exchange field
that is determined solely by the properties of the EuS/Al interface
and the normal-state properties of the Al films, such as their
thickness and conductance.  Consequently, the exchange field should,
in fact, be insensitive to the phase of the Al film, which is the
case for the EuS/Al bilayers in this study.  Preliminary measurements show that a very similar, applied field-dependent, exchange field arises in the Be component of EuS/Be bilayers \cite{XA}.  In fact, this exchange field is evident even in samples with Be layers of sufficiently high resistance so as to be in the non-superconducting correlated insulator phase \cite{BDA}.

 All of the current theoretical models treat the
exchange field within the context of a superconducting ground state,
and none of them can account for the fact that the exchange field is
an intrinsic function of the applied field.  If the underlying
mechanism of this field dependence can be determined, then one would
hope that the mechanism could be exploited in order to control the
magnitude of the exchange field with substantially smaller external
fields than used in this study.  Or, perhaps, one may be able to modulate the interface exchange
coupling with an external electric field via a gate.
In either case, the strategy is to use a small external field to control a large exchange field in order to realize a device, such as superconducting switch or a tunable polarized current source  \cite{Giazotto}.

\acknowledgments

We gratefully acknowledge enlightening discussions with Dan Sheehy and Ilya Vekhter.  PWA acknowledges the support if the DOE under
Grant No.\ DE-FG02-07ER46420 and GC the support of Yale University.  SS acknowledges the support of NSF under grant No. DMR-0545728.

\end{document}